\documentclass[%
reprint,
superscriptaddress,
showpacs,preprintnumbers,
amsmath,amssymb,
aps,
longbibliography,
pra,
floatfix,
]{revtex4-2}
\usepackage[utf8]{inputenc}
\usepackage[thinspace]{SIunits}
\usepackage{xspace}
\usepackage{graphicx}
\usepackage{dcolumn}
\usepackage{bm}
\usepackage{color}
\usepackage{soul}
\usepackage{multirow}
\usepackage{textgreek}
\usepackage{placeins}
\usepackage{adjustbox}
\usepackage{hyperref}
\usepackage{amssymb}
\usepackage{afterpage}
\usepackage{soul}
\usepackage{subfig}
\usepackage{caption}
\def\adda#1{}

\def\maa#1{}

\hypersetup{
    unicode = false,          
    pdftoolbar = true,        
    pdfmenubar = true,        
    pdffitwindow = false,     
    pdfstartview = {FitH},    
    pdftitle = {Raman spectroscopy of $2H-TaSe_{2-x}S_x$},    
    pdfnewwindow = true,      
    colorlinks = true,       
    linkcolor = blue,          
    citecolor = blue,        
    filecolor = blue,      
    urlcolor = blue,       
    plainpages = false,        
}

\makeatletter
\DeclareRobustCommand{\textsupsub}[2]{{%
  \m@th\ensuremath{%
    ^{\mbox{\fontsize\sf@size\z@#1}}%
    _{\mbox{\fontsize\sf@size\z@#2}}%
  }%
}}
\makeatother

\newcommand{\TSe}{\mbox{TaSe$_2$}\xspace}
\newcommand{\TS}{\mbox{TaS$_2$}\xspace}

\newcommand{\Alg}{\texorpdfstring{\ensuremath{A_{1g}}\xspace}{A1g}}

\newcommand{\Elg}{\texorpdfstring{\ensuremath{E_{1g}}\xspace}{Eg}}
\newcommand{\Ezg}{\texorpdfstring{\ensuremath{E_{2g}^2}\xspace}{Eg}}

\usepackage{array}
\newcolumntype{L}[1]{>{\raggedright\let\newline\\\arraybackslash\hspace{0pt}}m{#1}}

\newcommand{\wn}{\ensuremath{\rm cm^{-1}}\xspace}

\begin{document}
\title{Competition of disorder and electron-phonon coupling in \textit{2H}--TaSe{\boldmath{$_{2-x}$}}S{\boldmath{$_x$}} (0\ensuremath{\le}\boldmath{$x$}\ensuremath{\le}2) as evidenced by Raman spectroscopy}
\date{\today}
\author{J. Blagojevi\'{c}*}
\affiliation{Institute of Physics Belgrade, University of Belgrade, Pregrevica 118, 11080 Belgrade, Serbia}
\author{S. Djurdji\'{c} Mijin*}
\affiliation{Institute of Physics Belgrade, University of Belgrade, Pregrevica 118, 11080 Belgrade, Serbia}
\affiliation{Departamento de Física de Materiales, Facultad de Ciencias, Universidad Autónoma de Madrid, 28049 Madrid, Spain}
\author{J. Bekaert}
\affiliation{Department of Physics \& NANOlab Center of Excellence, University of Antwerp, Groenenborgerlaan 171, B-2020 Antwerp, Belgium}
\author{M. Opa\v{c}i\'{c}}
\affiliation{Institute of Physics Belgrade, University of Belgrade, Pregrevica 118, 11080 Belgrade, Serbia}
\author{Y. Liu}
\altaffiliation{Present address: Los Alamos National Laboratory, Los Alamos, New Mexico 87545, USA}
\affiliation{Condensed Matter Physics and Material Science Department, Brookhaven National Laboratory, Upton, NY 11973-5000, USA}
\author{M. V. Milo\v{s}evi\'{c}}
\affiliation{Department of Physics \& NANOlab Center of Excellence, University of Antwerp, Groenenborgerlaan 171, B-2020 Antwerp, Belgium}
\author{C. Petrovi\'{c}}
\affiliation{Condensed Matter Physics and Material Science Department, Brookhaven National Laboratory, Upton, NY 11973-5000, USA}
\affiliation{Shanghai Advanced Research in Physical Sciences (SHARPS), Shanghai 201203, China}
\affiliation{Department of Nuclear and Plasma Physics, Vinca Institute of Nuclear Sciences, University of Belgrade, Belgrade 11001, Serbia}

\author{Z. V.~Popovi\'{c}}
\affiliation{Serbian Academy of Sciences and Arts, Knez Mihailova 35, 11000 Belgrade, Serbia}
\author{N.~Lazarevi\'{c}}
\affiliation{Institute of Physics Belgrade, University of Belgrade, Pregrevica 118, 11080 Belgrade, Serbia}

\begin{abstract}
The vibrational properties of \textit{2H}-{TaSe$_{2-x}$S$_x$} (0\ensuremath{\le}$x$\ensuremath{\le}2) single crystals were probed using Raman spectroscopy and density functional theory calculations. The end members revealed two out of four symmetry-predicted Raman active modes, together with the pronounced two-phonon structure, attributable to the enhanced electron-phonon coupling. Additional peaks become observable due to crystallographic disorder for the doped samples. The evolution of the \Ezg mode Fano parameter reveals that the disorder has weak impact on electron-phonon coupling, which is also supported by the persistence of two-phonon structure in doped samples. As such, this research provides thorough insights into the lattice properties, the effects of crystallographic disorder on Raman spectra, and the interplay of this disorder with the electron-phonon coupling in \textit{2H}-{TaSe$_{2-x}$S$_x$} compounds.
\end{abstract}

\pacs{%
}
\maketitle

\section{Introduction}
Transition metal dichalcogenides, a well studied family of quasi-2D materials, have attracted considerable attention in the recent years due to their rich phase diagrams, thickness-dependent transport, unique optical properties and collective electron phenomena (e.g. charge density waves and superconductivity) \cite{doi:10.1126/science.aac9439, doi:10.1080/00018736900101307, PhysRevLett.86.4382, D1CS00236H, articlewang}. Contrary to what previous theoretical studies suggested, experimental results have demonstrated the possible coexistence of superconductivity and CDW \cite{HESS1991422, doi:10.1080/00018737500101391}. Given that these phenomena arise at experimentally accessible temperatures, transition metal dichalcogenides represent ideal candidates for their investigation \cite{SHI20181, koley2020charge}.\par
Previous experimental research has shown that at room temperature both the \textit{2H}-$\mathrm{TaS_2}$ and the \textit{2H}-$\mathrm{TaSe_2}$ crystallize into the hexagonal structure, described by the space group $P6_3/mmc$ ($D_{6h}$) \cite{fazni1, fazni2}. The opulent phase diagram of \textit{2H}-$\mathrm{TaSe_2}$ includes numerous charge density wave (CDW) phases at high temperatures -- the incommensurate CDW (ICCDW) phase at $T_{1C}$ = 122 K, the single commensurate CDW (SCCDW) phase in the temperature range from $T_{\downarrow 2C}$ = 112 K to $T_{\uparrow 2C}$ = 90 K, and the triply commensurate (TCCDW) phase at $T_{3C}$ = 90 K \cite{fazni1, fazni2, fazni3, fazni5}. As for \textit{2H}-$\mathrm{TaS_2}$, transition from the normal (metallic) to the ICDW phase occurs at $T_{CDW}$ = 78 K \cite{fazni4}. These materials exhibit unusually large Raman two-phonon scattering cross section, often correlated with the existence of CDW phase \cite{fazni2, doi:10.1063/5.0051112, fazni4, behera1986theory}. Two-phonon feature in \textit{2H}-$\mathrm{TaS_2}$ was attributed to second-order scattering of acoustic and quasi-acoustic modes near the $q_{CDW}$ $\cong$ $\frac{2}{3}{\Gamma}$M \cite{PhysRevB.99.245144}.\par
The latest experimental results indicate that the substitution of Se atoms with S atoms leads to a weak double dome evolution of the superconducting critical temperature \textit{$T_{SC}$}. The ${T_{SC}}$ dependence coincides with the evolution of crystallographic disorder, suggesting that the crystalline disorder favors superconductivity while suppressing the CDW phase \cite{disordercdw}. Similarly, other types of disorder have been demonstrated to engender enhanced superconductivity in this system, such as by etching nanopores in monolayer $\mathrm{TaS_2}$ sheets \cite{Peng2018}, or by spontaneous filling of vacant sulfur sites by oxygen in few-layer \textit{2H}--$\mathrm{TaS_2}$ samples \cite{Bekaert2020}. The latter yields another example of isoelectronic substitution in this system, similar to the exchange of sulfur and selenium. In addition, first-principles calculations have revealed that the electron-phonon coupling within the $\mathrm{TaS_2}$ sheets is drastically boosted (by oxygenation up to 80\%) \cite{Bekaert2020}, providing yet another pathway to enhance superconductivity by doping, in addition to the suppression of the CDW state.\par
In this work, we present a Raman spectroscopy study of \textit{2H}-{TaSe$_{2-x}$S$_x$} (0\ensuremath{\le}$x$\ensuremath{\le}2) alloys. Obtained experimental results were found to be in good agreement with the density functional theory (DFT) calculations. The experimental Raman spectra of the end compounds host two out of the four symmetry-predicted Raman active modes. Additionally, a low-intensity overtone peak $O_1$ obeying pure \Alg selection rules can be observed only in the spectra of \textit{2H}--TaS$_2$. We have revisited the origin of the two-phonon structure and concluded that it results from enhanced electron-phonon coupling within phonon branches around the $M$ and $L$-points. In the spectra of doped samples additional peak and a dynamic evolution of two-phonon structure are observed due to crystallographic disorder. Our analysis of the inverse Fano parameter 1/$|q|$ of \Ezg mode indicates that crystallographic disorder has weak effect on the electron-phonon coupling. 

\section{Experimental and Computational Details}

The Raman experiment was performed using a Tri Vista 557 spectrometer with a 1800/1800/2400 grooves/mm diffraction grating combination in a backscattering configuration. As an excitation source, the 514 nm line of a Coherent Ar$^+$/Kr$^+$ ion laser was used. The direction  of the incident (scattered) light coincides with the crystallographic $c$ axis. Laser beam focusing was achieved through a microscope objective with 50$\times$ magnification. During the measurements the samples were placed inside of a KONTI CryoVac continuous helium flow cryostat with 0.5 mm thick window. All samples were cleaved in the air before being placed into the cryostat. The obtained Raman spectra were corrected by the Bose factor. The spectrometer resolution is comparable to the Gaussian width of 1 \wn.\par
\begin{figure}[!htbp]
\centering
\includegraphics[width = 85mm]{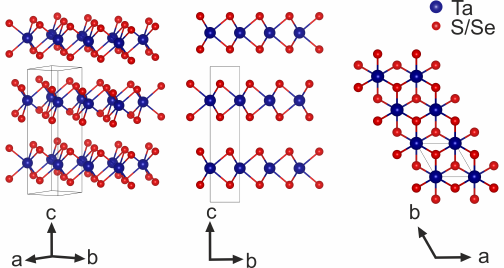}
 \caption{Schematic representation of crystal structure of \textit{2H}--$\mathrm{TaSe_2}$ (\textit{2H}--$\mathrm{TaS_2}$) in various orientations.}
\label{Figure1}
\end{figure}
Next, we have performed density functional theory (DFT) calculations as implemented in the ABINIT package \cite{GONZE2020107042}. We have used the Perdew-Burke-Ernzerhof (PBE) functional, an energy cutoff of 50 Ha for the plane-wave basis, and included spin-orbit coupling by means of fully relativistic Goedecker pseudo-potentials \cite{PhysRevB.54.1703, Krack2005}, where Ta-5d$^3$6s$^2$, S-3s$^2$3p$^4$, and Se-4s$^2$4p$^4$ states are treated as valence electrons. The crystal structure was relaxed so that forces on each atom were below 0.05 meV/\AA~and the total stress on the unit cell below 0.1 bar. This relaxation yields lattice parameters $a = 3.39$ \AA, $c = 14.00$ \AA~for TaS$_2$, and $a = 3.51$ \AA, $c = 14.37$ \AA~for TaSe$_2$. Subsequently, the phonons and the electron-phonon coupling (EPC) were obtained from density functional perturbation theory (DFPT) calculations, also within ABINIT \cite{PhysRevLett.68.3603}. Here, we have used a $16 \times 16 \times 6$ $\textbf{k}$-point grid for the electron wave vectors and an $8 \times 8 \times 3$ $\textbf{q}$-point grid for the phonon wave vectors. For the electronic occupation we employed Fermi-Dirac smearing with broadening factor $\sigma_\mathrm{FD} = 0.01$ Ha (sufficiently high to exclude unstable phonon modes related to the low-temperature CDW phases).

\section{Results and Discussion}


\subsection{\textit{2H}--$\mathbf{TaSe_2}$ and \textit{2H}--$\mathbf{TaS_2}$}

All single crystal alloys of \textit{2H}--{TaSe$_{2-x}$S$_x$}  (0\ensuremath{\le}$x$\ensuremath{\le}2) crystallize into \textit{P6$_{3}$/mmc} crystal structure (Fig.~\ref{Figure1}) \cite{fazni1, fazni2}. Wyckoff positions of atoms and their contributions to the $\Gamma$-point phonons together with corresponding Raman tensors for the two end compounds are listed in Table~\ref{ref:Table1}.
\begin{table}[!htbp]
\caption{Wyckoff positions of atoms and their contributions to the $\Gamma$-point Raman active phonons for the ${P6_{3}/mmc}$ space group of \textit{2H}--$\TSe$ (and \textit{2H}--$\TS$) together with the corresponding Raman tensors.}
\label{ref:Table1}
\begin{ruledtabular}
\centering
\resizebox{\linewidth}{!}{%
\centering
\begin{tabular}{ccc}
\multicolumn{3}{c} {Space group: ${P6_{3}/mmc\;(194)}$} \\ \cline{1-3} \\[-2mm]

Atoms &  \multicolumn{2}{c} {Irreducible representations} \\ \cline{1-1} \cline{2-3} \\[-2mm]

Ta  ($2b$) & \multicolumn{2}{c} {$E_{2g}$} \\[1mm]

Se/S ($4f$) & \multicolumn{2}{c} {$A_{1g}+ E_{1g}+ E_{2g}$} \\[1mm]

\cline{1-3}\\[-2mm]
\multicolumn{3}{c} {Raman tensors} \\ \cline{1-3} \\[-2mm]

\multicolumn{3}{c}{
$\Alg$  = $\begin{pmatrix}
a&0 &0 \\
 0&a& 0\\
 0& 0&b\\
\end{pmatrix}
$}
\\ [5mm]
\multicolumn{3}{c}{$
{}^1\Elg  = \begin{pmatrix}
0&0 &0 \\
0& 0& c\\
0 & c &0 \\ \end{pmatrix}
\;
{}^2\Elg  = \begin{pmatrix}
0& 0&-c\\
0& 0&0\\
-c&0 &0 \\
\end{pmatrix}
$}
\\ [5mm]
\multicolumn{3}{c}{$
{}^1\Ezg  = \begin{pmatrix}
d&0&0\\
0&-d&0\\
0&0&0 \\ \end{pmatrix}
\;
{}^2\Ezg  = \begin{pmatrix}
0&-d&0\\
-d&0&0\\
0&0& 0\\
\end{pmatrix}
$}

\end{tabular}}
\end{ruledtabular}
\end{table}

\begin{figure*}[!htbp]
\centering{
	\includegraphics[width = 170mm]{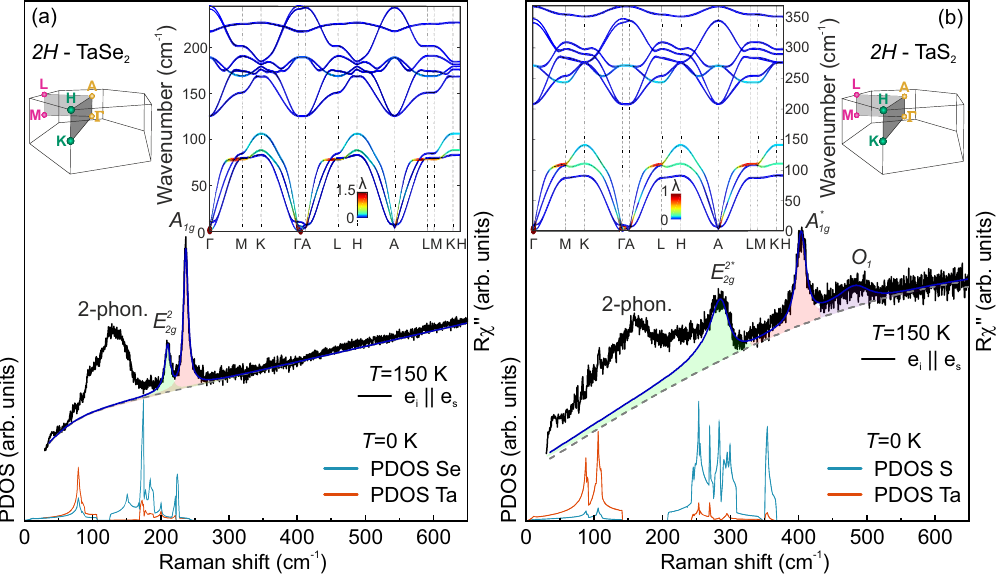}
	\label{Figure2}
	}
\caption{Raman response of \textit{2H}--TaSe$_2$ (a) and \textit{2H}--TaS$_2$ (b) for parallel polarization configuration at $T$ = 150 K. The blue solid line represents the cumulative fit consisting of electronic continuum (dashed line) and phonon modes. The electronic continuum was modeled using simplified approach as described in \cite{mialitsin2011}, utilizing the following function: $R^{''}_{\chi}(\omega, T) = a(T)\cdot\tanh{\frac{\omega}{c(T)}}+b(T)\cdot{\frac{\omega}{c(T)}}$. Here $a(T)$, $b(T)$ and $c(T)$ are temperature dependent parameters determined for each sample. Phonon modes were fitted using Fano line profile. Parts of spectra not covered by cumulative fit were identified to originate from two-phonon scattering. The insets shows the calculated phonon dispersion of both pure samples in the normal phase, with the electron-phonon coupling ($\lambda$) indicated through the color scale. The contributions of the Ta (orange) and Se/S atoms (blue) to the calculated PDOS are shown below. First Brillouin zone with indicated points and lines of high symmetry is also shown in the inset.}
\label{Figure2}
\end{figure*}
In total, the symmetry analysis predicts four Raman-active modes (\Alg + \Elg + 2$E_{2g}$) for the two end compounds of \textit{2H}--{TaSe$_{2-x}$S$_x$}. According to the Raman tensors presented in Table~\ref{ref:Table1}, \Alg modes can be observed only in parallel polarization configuration, whereas $E_{2g}$ modes can be observed in spectra measured both in parallel and crossed polarization configurations. For our backscattering configuration, where laser beam is focused along the $c$-axis onto $ab$ plane, the \Elg mode is unobservable.\par
Raman spectra of the two end compounds \textit{2H}--TaS$_2$ and \textit{2H}--TaSe$_2$ measured at $T$ = 150 K are presented in Fig.~\ref{Figure2}. This particular temperature has been chosen because it is significantly higher than the critical temperature of the CDW phase transitions. 
The Raman spectra of the pristine samples exhibit two prominent peaks, assigned as \Alg and \Ezg symmetry modes, accompanied by a broad structure and an additional peak in the case of \textit{2H}-TaS$_2$. For clarity, Raman-active modes observed in \textit{2H}--$\TS$ spectra are marked with asterisk. Theoretically predicted modes are first analyzed, after which we address the origin and nature of additional peak and broad structure.\par
\begin{table}[!htbp]
\caption{Phonon symmetries and phonon energies (in units of $\wn$) at $\Gamma$ and their degeneracies for the ${P6_{3}/mmc}$ structure of \textit{2H}--$\TS$ and \textit{2H}--$\TSe$. The experimental values were determined at 300\,K with experimental uncertainty of 0.3 \wn. The DFPT calculations were performed at zero temperature. In case of the lowest two modes, `S' stands for shear and `LB' for layer breathing.}
\label{ref:Table2}
\begin{ruledtabular}
\centering
\resizebox{\columnwidth}{!}{%
\centering
\begin{tabular}{cccccc}
\\[-2.5mm]
& & \multicolumn{2}{c}{\textit{2H}--$\TS$} & \multicolumn{2}{c}{\textit{2H}--$\TSe$} \\[1mm] 

 \cline{3-4} \cline{5-6} \\[-1mm]

Symm. & Deg. & Calc. & Exp. & Calc. & Exp.\\[0.5mm] 

\cline{1-2} \cline{3-4} \cline{5-6} \\[-1mm]

$E_{2g}^1$ (S) & 2 & 4.9 & & 5.5 & \\ [0.5mm]

$B_{1g}^1$ (LB) & 1 & 10.4 & & 10.7 & \\ [0.5mm]

$E_{2u}$ & 2 & 208.0 & & 125.4 & \\ [0.5mm]

$E_{1g}$ & 2 & 208.1 & & 125.4 & \\ [0.5mm]

$E_{2g}^2$ & 2 & 271.3 & 288.8 & 189.1 & 207.7\\[0.5mm]  

$E_{1u}$ & 2 & 271.3 & & 189.2 & \\ [0.5mm]

$A_{2u}$ & 1 & 341.6 & & 241.8 & \\ [0.5mm]

$B_{1g}^2$ & 1 & 348.0 &  & 245.1 & \\ [0.5mm] 

$B_{1u}$ & 1 & 368.0 & & 216.9 & \\ [0.5mm]

\Alg & 1 & 368.5 & 402.9  & 217.7 & 234.7 \\
\end{tabular}}
\end{ruledtabular}
\end{table}
The Raman responses of both samples were analyzed with a cumulative fit to include electronic continuum and discrete single-phonon excitations. The Breit-Wigner-Fano profile was used to model single-phonon excitations, due to visible asymmetry of the \Ezg modes. For simplicity, the \Alg modes were fitted with constant Fano parameter $|q| = 50$, as they are highly symmetric. The phonon energies obtained in this way, alongside the calculated ones are listed in Table~\ref{ref:Table2}. The lower frequency $E_{2g}^1$ modes are related to shear (S) movements of the layers, with frequencies of $\sim 5$ cm$^{-1}$ for both compounds, whereas $B_{1g}^1$ modes represent the layer breathing (LB) and are expected at frequencies $\sim 10$ cm$^{-1}$.\par
As can be seen in Table~\ref{ref:Table2}, the discrepancy in experimental and theoretical phonon energy is less than 10 \% for all observed modes. The $E_{2g}^2$ modes were not observed in the investigated energy region in accordance to our numerical calculations and are not in the focus of this study. The phonon dispersions, obtained from the DFPT calculations, are shown in the insets of Fig.~\ref{Figure2}, together with the calculated values of the EPC constant $\lambda$. These indicate the presence of heightened EPC in \Ezg scattering channel, giving rise to asymmetric line profiles. On the other hand in the \Alg channel no significant EPC was found, thus they should have symmetric profiles, as we have observed.\par
In addition to the observed Raman-active modes, the spectra of \textit{2H}--TaS$_2$ in the parallel polarization configuration host the $O_1$ peak, with energy at about $\sim$ 490 \wn.  This unexpected peak is not predicted by theoretical calculations and cannot be explained in terms of the first-order Raman scattering as its energy is well beyond the phonon energy range (see Fig.~\ref{Figure2} (b)). Although the $O_1$ peak is not a result of single-phonon excitation, it was included in the cumulative fit in order to track its energy. Previously reported large two-phonon scattering cross section \cite{fazni2, doi:10.1063/5.0051112, PhysRevB.99.245144, fazni4, behera1986theory} indicate that this peak is most likely the result of a second-order Raman scattering, which might occur due to defect and/or enhanced electron-phonon coupling. Assuming that the second-order Raman peak $O_1$ becomes observable due to existing defects in the measured crystals, additional first-order peaks at energies with the highest phonon density of states (PDOS) values should also arise in the Raman spectra \cite{Baum2018_PRB97_054306}. However, no such peaks were observed. Considering that one of the PDOS maxima can be located at the energies that correspond to the half energy of $O_1$ peak, we believe that this peak is an overtone in nature and is observable due to enhanced EPC. To determine the validity of this assumption, the phonon dispersion curves and electron-phonon coupling constant $\lambda$ (depicted as inset in Fig.~\ref{Figure2} (b)) were further examined. If the $O_1$ peak is observable due to electron-phonon coupling, we would expect that optical phonon branches around energies $\omega_{O_1}/2$ express high values of $\lambda$. As it is indicated in the inset of Fig.~\ref{Figure2} (b), several optical branches along the lines of high symmetry, at energies just below 250 \wn meet this condition. Thus we argue in favor of the assumption that the $O_1$ peak most likely originates from the heightened electron-phonon coupling.\par
The broad structures in the spectra of \textit{2H}--TaSe$_2$ and \textit{2H}--TaS$_2$, with energies at about $\sim$ 130 \wn and $\sim$ 160 \wn, respectively, are centered in the gap of the theoretical PDOS. That means they cannot be a product of a first-order Raman scattering. Considering previous discussion on the origin of the $O_1$ peak in \textit{2H}--TaS$_2$ spectra and applying similar analysis, these structures can likewise be explained as a two-phonon processes, predominantly overtones in nature but also with combinations contribution (as can be deduced from Fig. \ref{FigureA3}), also originating from the enhanced electron-phonon coupling. As shown in Fig.~\ref{Figure2}, peaks in PDOS values can be found in the $\omega_{2phon}/2$ energy ranges. By examining the phonon dispersion curves of \textit{2H}--TaS$_2$ and \textit{2H}--TaSe$_2$ (insets in Fig.~\ref{Figure2} (a) and (b), respectively), several phonon branches with enhanced electron-phonon coupling are in the appropriate energy range for the two-phonon process. In the case of \textit{2H}--TaS$_2$, these branches fall within the energy range of 100 - 150 \wn, while for \textit{2H}--TaSe$_2$, they lie in the range of 50 - 100 \wn.
It was reported in a previous study \cite{PhysRevB.99.245144} that the origin of two-phonon structure in \textit{2H}--TaS$_2$ is in the vicinity of the CDW wave vector $q_{CDW}\cong\frac{2}{3}\Gamma$M, where indeed strong electron-phonon coupling exists. We argue that the origin of large scattering cross section for the two-phonon process is enhanced electron-phonon coupling in a much larger phase space that is required to reproduce the high Raman cross-section of the observed structures.
The phonon branches along (and in the vicinity) the lines of high symmetry: $\Gamma-M$, $A-L$, and $L-M$ (illustrated in Fig.~\ref{Figure2}), demonstrate the highest values of the electron-phonon coupling constant, and we believe that these regions of phase space contribute to the two-phonon process. These branches also closely coincide with energy levels that correspond to $\omega_{2phon}/2$, as we anticipated since the major contribution lies in the \Alg symmetry corresponding to the overtones (see Appendix \ref{A1}). This provides a strong argument to support our assumptions, regarding the nature and the origin of the two-phonon structures. Significantly higher intensity of the two-phonon structure in \textit{2H}--TaS$_2$, compared to the $O_1$ peak, can be attributed to the stronger electron-phonon coupling observed in the corresponding phonon branches, as supported by our DFT calculations (Fig.~\ref{Figure2} (b)).

\subsection{\textit{2H}--TaSe{\boldmath{$_{2-x}$}}S{\boldmath{$_x$}} (0\ensuremath{\le}\boldmath{$x$}\ensuremath{\le}2)}
\begin{figure}[!tp]
\centering
\includegraphics[width = 85mm]{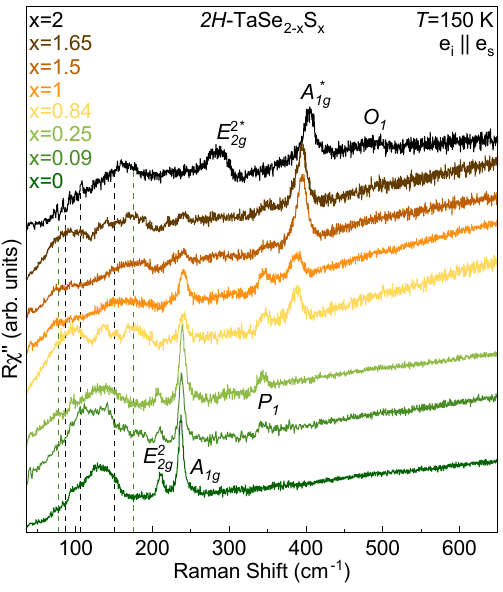}
 \caption{Raman spectra of \textit{2H}--{TaSe$_{2-x}$S$_x$} measured at 150 K in parallel polarization configuration. Additionally to the \Alg, $\Alg^{*}$, \Ezg and $E_{2g}^{2*}$ modes, one peak assigned as $P_1$ and PDOS projection onto two-phonon structure arise in the spectra of doped materials. With dashed lines are marked places of PDOS maxima in both pure samples.}
\label{Figure3}
\end{figure}

Raman spectra of the \textit{2H}--{TaSe$_{2-x}$S$_x$} (0\ensuremath{\le}{$x$}\ensuremath{\le}2) single crystals measured in parallel polarization configuration at $T$ = 150 K are presented in Fig.~{\ref{Figure3}}. The two modes that are present in the spectra of pristine \textit{2H}--$\mathrm{TaSe_{2}}$ - \Alg and \Ezg, can also be observed in the spectra of the doped samples. 
The \Alg symmetry mode persists in the spectra up to \textit{2H}--$\mathrm{TaSeS}$. With further increase in the value of $x$ we cannot be certain that the existing feature in spectra is \Alg mode, because of low Se content and possible \textit{2H}--$\mathrm{TaS_{2}}$ PDOS contributions. On the other hand \Ezg vanishes when sulfur content reaches the value of $x$ = 0.84. The $\Alg^{1*}$ and $E_{2g}^{2*}$ modes can last be found in the spectra of \textit{2H}--$\mathrm{TaSe_{1.16}S_{0.84}}$ and \textit{2H}--$\mathrm{TaSe_{0.5}S_{1.5}}$, respectively.
Taking into consideration that \Ezg and $E_{2g}^{2*}$ are modes arising from the out of phase in-plane oscillations of Ta and Se/S atoms, the dependence on sulfur content $x$ exhibits a discontinuous behavior. For the \Alg mode (out-of-plane movement of Se/S atoms) we observe continuous change with doping alongside simultaneous coexistence of \Alg and $\Alg^*$ peaks in samples where Se/S concentration ratio is around unity. This dichotomous behavior of phonon modes is quite intriguing and somewhat unexpected, although similar case is reported in \cite{lazarevic2022evolution}. 
 In the specific case of \textit{2H}--$\mathrm{TaSeS}$, at the nanoscale level, localized regions may emerge where either Se or S atoms predominate. Although these regions possess a sufficient size to generate detectable signals, they are significantly smaller in comparison to the laser spot ($\sim$ 5 {\textmu}m$^2$), resulting in the inclusion of a considerable number of such clusters of atoms within the laser's spatial coverage. Consequently, the resulting Raman spectrum displays discernible contributions from both elements. The concentrations at which the \Ezg mode vanishes directly correlate with disorder reaching its maximum value. Given that doping modifies the bond lengths of Ta-S and Ta-Se atoms, which, in conjunction with the accompanying lattice disorder, has a detrimental effect on the \Ezg mode.\par
\begin{figure}[!tp]
\centering
\includegraphics[width = 85mm]{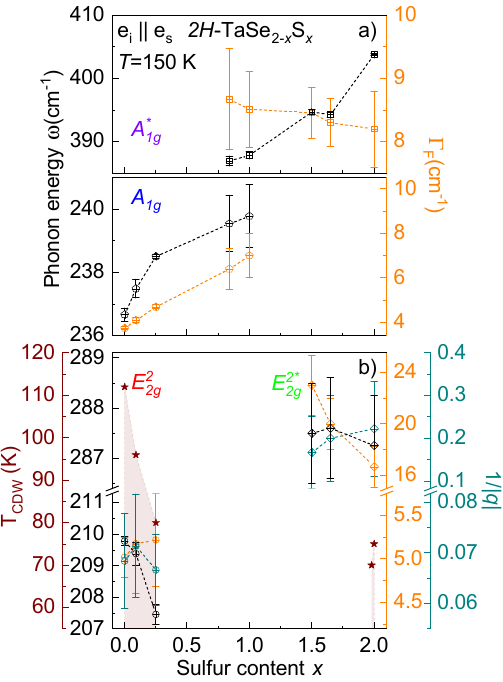}
\caption{(a) The evolution of phonon energies (black) and line widths (orange) of the \Alg, $\Alg^{1*}$  Raman-active modes. (b) The evolution of phonon energies, line widths  and inverse value of Fano parameter 1/$|q|$ (teal) of the \Ezg and $E_{2g}^{2*}$ modes with the sulfur content $x$. $T_{CDW}$ (dark red) dependence on the sulfur content $x$, together with the electronic phase diagram (shaded red) of \textit{2H}--{TaSe$_{2-x}$S$_x$} reflect the evolution of crystallographic disorder \cite{disordercdw}.}
\label{Figure4}
\end{figure}

The gradual substitution of the selenium atoms with sulfur in the \textit{2H}--$\TSe$ crystals introduces crystallographic disorder, giving rise to new scattering channels that result in additional peak observed in the Raman spectra of doped materials. This new peak with energy at about 342 \wn, assigned as $P_1$ is observed exclusively in the doped spectra measured in parallel polarization configuration, therefore obeying \Alg symmetry rules. It is presumably an overtone mode, as its energy is almost exactly double that of the location where the highest PDOS value of the \textit{2H}--$\TSe$ is situated. To further investigate the effects of crystallographic disorder, we have inspected the evolution of phonon parameters with the sulfur content $x$, depicted in Fig.~\ref{Figure4}. As in the case of pure samples, all phonon lines were fitted using Fano profiles. No asymmetry was observed for the \Alg modes and they were fitted with the value of Fano parameter $q$ being fixed at 50. All modes but \Ezg harden with the increase in the sulfur concentration in the measured crystals. 
Given the difference in the atomic mass of Se and S and a reduction of the unit cell volume \cite{disordercdw}, one would expect that the \Ezg mode also hardens with increasing $x$ \cite{PhysRevB.58.43, Charrier-Cougoulic, PhysRevB.81.024304}. The unexpected behavior of \Ezg mode might be attributed to the enhanced electron-phonon coupling, which potentially overcompensates previously mentioned effects that lead to the hardening of the mode. All Raman modes broaden due to increased crystallographic disorder, which is to be expected. Considering that the Fano profiles are used to describe the line shape originating from coupling between phonon and electronic continuum, and that the Fano parameter depends on the interaction strength between the phonon and the continuum, the inverse value of Fano parameter 1/$|q|$ can be used as a direct measure of the strength of electron-phonon coupling \cite{PhysRev.124.1866, Sathe_2016, PhysRevB.62.4142, ephon}.
The interference of two-phonon structure in the fitting procedure produces large uncertainty (see Fig.~\ref{Figure4} (b)), thus we cannot claim whether the EPC slightly decreases or increases, but can only argue that it persists through growing disorder. This result falls within the range of the potential increase in disorder-induced EPC \cite{disordercdw}.\par
Crystallographic disorder also has a significant impact on the evolution of two-phonon structure. As mentioned, the broad structure in pristine \textit{2H}--$\TSe$, originating from the two-phonon process, can be found in the spectra centered at around 130 \wn. Changing the sulfur content to $x$ = 0.09, the structure undergoes a further broadening accompanied by pronounced evolution of the shoulder at around $\sim$ 110 \wn, likely due to more pronounced evolution of the  EPC in the related phonon branches. Raising the sulfur content to 0.25 and thus further increasing disorder restores the structure to a solitary broad peak. However, the most drastic change in the evolution of the spectra is observed for $x$ = 0.84. In this case there are now four first order peaks superimposed onto a broad two-phonon structure. As discussed previously, the additional peaks might become observable in Raman experiment due to defect scattering and are related to the regions where the PDOS reaches its maximum: at $\sim$ 90 \wn and $\sim$ 110 \wn for \textit{2H}--$\TS$, at $\sim$ 80 \wn and $\sim$ 175 \wn for \textit{2H}--$\TSe$ (see Fig.~\ref{Figure2}). These correspond rather well with the energies of the newly observed phonon modes.
Raising the concentration of sulfur to $x$ = 1 leads to another drastic change in the two-phonon structure, as it is now wholly diluted and has two broad peaks at around $\sim$ 80 \wn and $\sim$ 160 \wn. In the spectra with sulfur content $x$ = 1.65, structure resembles the structure observed in the the spectrum of the compound with sulfur content $x$ = 0.84, albeit of less clarity among peaks. From these observations, we can see that the primary impact of disorder on the two-phonon structure is manifested by the broadening of peaks and the projection of the PDOS. Furthermore, the persistence of the two-phonon structure across doped samples suggests that disorder does not hider the electron-phonon coupling. This corollary aligns with previous observations regarding the $1/|q|$ parameter of the \Ezg mode.\par
Our results reveal the origin of two-phonon structures in the \textit{2H}--TaSe$_{2-x}$S$_x$ compounds. It is also evident that, although disorder does not directly influence the strength of the EPC (in the range from $x$ = 0.25 to $x$ = 1.65), it affects the phenomena that share EPC as a seemingly common origin differently. While the CDW is destroyed, SC and the two-phonon structure survive, suggesting that strong EPC alone is insufficient to sustain the CDW in a disordered lattice. It is interesting to note that these materials simultaneously and independently experience the effects of disorder and EPC.
\section{Conclusions}

In this study, we conducted a Raman scattering analysis of \textit{2H}-{TaSe$_{2-x}$S$_x$} (0\ensuremath{\leq}{$x$}\ensuremath{\leq}2) alloys. The Raman spectra of the end compounds host two out of the three symmetry-expected Raman active modes for backscattering configuration and an additional peak $O_1$ of \Alg symmetry, present only in the \textit{2H}--TaS$_2$ spectra. This $O_1$ peak is a result of second-order Raman scattering, overtone in nature, observable due to the prominent EPC, characteristic for CDW materials. The broad two-phonon structures observed in pristine samples were linked to enhanced EPC in the corresponding phonon branches, as obtained from our first-principles calculations. The gradual substitution of Se atoms with S atoms results in crystallographic disorder, introducing new scattering channels -- in the form of an additional peak $P_1$ of \Alg symmetry, as well as PDOS projection and broadening of all phonon modes. Dependence on sulfur content showed hardening of all modes, except for \Ezg mode. Softening of the \Ezg mode and its discontinuous dependence were attributed to strong EPC. Coexistence of \Alg peaks in intermediate doping levels arises probably due to nanoscale clusters comprised mainly of either Se or S atoms. In spectra where disorder reaches its maximum, four single-phonon peaks accompanied the background two-phonon structure, with energies corresponding to the region of PDOS maxima. The negligible influence of disorder on the EPC was supported by the continued presence of the two-phonon structure and the behavior of the $1/|q|$ parameter associated with the \Ezg mode. Thus, in the absence of changes in either electron-electron \cite{disordercdw} or EPCs, significant rise of $T_{SC}$
in doped alloys is only due to disorder-induced CDW suppression. Our findings provide insights into the intricate relationship between disorder and EPC, shedding light on their combined influence in governing the vibronic and collective electronic behavior of \textit{2H}--{TaSe$_{2-x}$S$_x$} compounds.
\begin{acknowledgments}
    
Authors J. Blagojevi\'{c} and S. Djurdji\'{c} Mijin contributed equally to the paper. The authors acknowledge funding provided by the Institute of Physics Belgrade through the grant by the Ministry of Science, Technological Development and Innovation of the Republic of Serbia, Project F-134 of the Serbian Academy of Sciences and Arts, the Science Fund of the Republic of Serbia, PROMIS, No. 6062656, StrainedFeSC, and by Research Foundation-Flanders (FWO). J. Bekaert acknowledges support as a Senior Postdoctoral Fellow of the FWO (Fellowship No. 12ZZ323N), and from the Erasmus+ program for staff mobility and training (KA107, 2018) for a research stay at the Institute of Physics Belgrade, during which part of this work was carried out. The computational resources and services used for the first-principles calculations in this work were provided by the VSC (Flemish Supercomputer Center), funded by the FWO and the Flemish Government - department EWI. Work at Brookhaven National Laboratory was supported by U.S. DOE, Office of Science, Office of Basic Energy Sciences under Contract No. DE-SC0012704.
\end{acknowledgments}
%
\clearpage
\appendix

\section{Mode assignation and nature of additional peaks}
\label{A1}
\begin{figure}[!htp]
\centering
\includegraphics[width = 85mm]{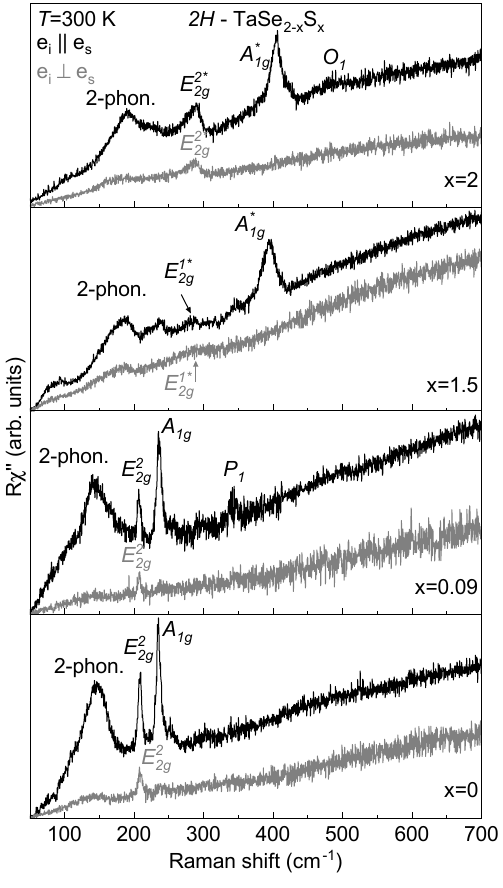}
\caption{Raman response of \textit{2H}--TaSe$_2$, \textit{2H}--TaSe$_{1.5}$S$_{0.5}$, \textit{2H}--TaSe$_{1.91}$S$_{0.09}$ and \textit{2H}--TaS$_2$ for parallel (black) and crossed (gray) polarization configurations at $T$ = 300 K. The possible overtone modes $O_1$ and $P_1$ are only present in parallel polarization configuration.}
\label{FigureA3}
\end{figure}
Raman spectra of all samples were measured in parallel and crossed polarization configurations in order to correctly assign theoretically predicted modes as well as the unexpected and additional peaks and structures. From the Raman spectra presented in Figure~\ref{FigureA3} theoretically predicted \Alg and \Ezg modes were easily identified. Given that the $O_1$ and $P_1$ peaks are only present in parallel polarization configuration, they obey pure \Alg symmetry rules. Two-phonon structure is present in both polarization configurations, but with significantly higher intensity in \Alg channel. Additional features in two-phonon structure in the spectra of \textit{2H}--TaSe$_{1.5}$S$_{0.5}$ are only present in parallel polarization configuration, thus they are possibly overtones in nature or single-phonon excitations of \Alg symmetry.

\section{Fitting details}

\begin{figure}[!hb]
\centering
\includegraphics[width = 85mm]{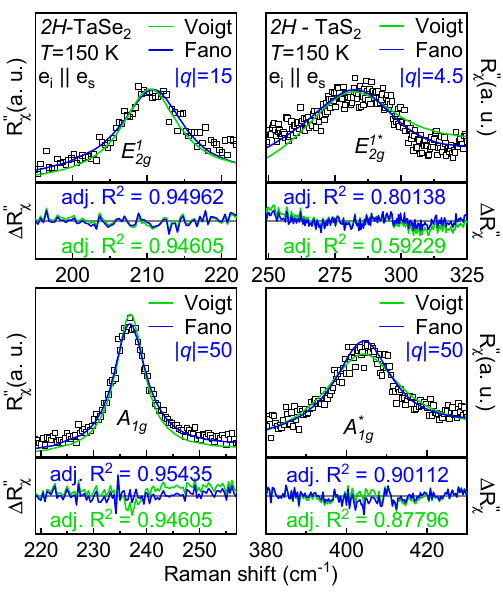}
\caption{Raman response as a function of the Raman shift. Quantitative analysis of the $\Ezg$, $E_{2g}^{2*}$, $\Alg$ and $\Alg^{*}$ modes at $T$ = 150 K. The blue and green solid lines represent Fano and Voigt profiles fitted to the experimental data, respectively. Experimental data is represented by open squares.}
\label{FigureA1}
\end{figure}

All peaks were fitted using both asymmetric Fano profiles and symmetric Voigt profiles. Since the line widths of the analyzed phonon modes were much greater than the resolution of spectrometer $\sigma$, the real width of the peaks could be obtained without the need to use Fano profiles convoluted with a Gaussian function, where $\Gamma_L$ = $\sigma$. The comparison between the obtained fits are presented in Figure~\ref{FigureA1}. As it can be seen, peaks $\Ezg$ and $E_{2g}^{2*}$ show clear asymmetric line shape, whereas the $\Alg$ symmetry peaks do not. Therefore, the latter were fitted with Fano parameter value $|q|$ being fixed at 50. Acquired phonon parameters as a function of sulfur content $x$ are presented in Figure~\ref{Figure4}.\par
\begin{figure}[!hp]
\centering
\includegraphics[width = 85mm]{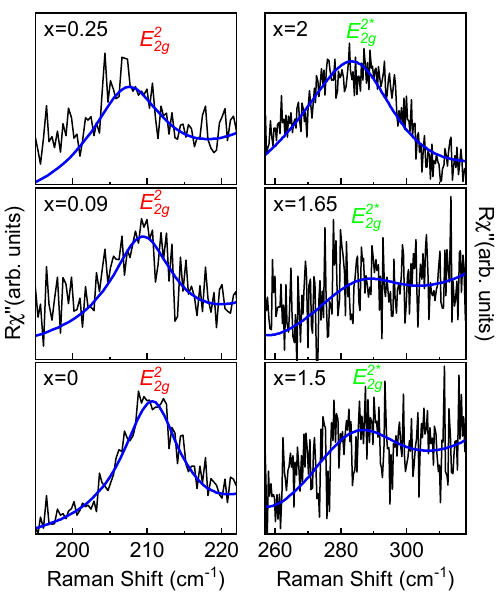}
\caption{Raman response as a function of the Raman shift. Quantitative analysis of the $\Ezg$ and $E_{2g}^{2*}$ modes for indicated sulfur content $x$ in the measured crystals. The blue lines represent Fano profiles fitted to the experimental data.}
\label{FigureA2}
\end{figure}

The increase in doping leads to the significant decrease in the relative intensity of $\Ezg$ Raman active modes, therefore the obtained phonon parameters have slightly higher error bar compared to the $\Alg$ modes. Fits obtained using Fano profiles in the energy range of peaks $\Ezg$ and $E_{2g}^{2*}$ are presented in Figure~{\ref{FigureA2}}. The two-phonon structure interferes with the lower energy side of the peaks in the spectra of all doped samples, contributing to the mentioned error. 

\end{document}